\documentclass[doublecol]{epl2} 

\usepackage[utf8]{inputenc}
\usepackage{graphicx}
\usepackage{physics, amsmath, amssymb, amsthm, units, dsfont, bm}
\usepackage{mathtools, amsfonts, mathrsfs, bbm}
\usepackage{xcolor}

\definecolor{dark-gray}{rgb}{.35,.55,.55}
\definecolor{dark-blue}{rgb}{.0,.0,.6}
\usepackage[colorlinks=true,linkcolor=dark-blue,citecolor=dark-blue,urlcolor=dark-blue]{hyperref}

\renewcommand{\bra}[1]{\langle #1\vert}
\renewcommand{\ket}[1]{\vert #1\rangle}
\renewcommand{\braket}[2]{\langle #1\vert #2\rangle}
\renewcommand{\ketbra}[2]{\vert #1 \rangle\! \langle #2\vert}

\title{Quantifying entanglement from the geometric perspective}

\author{Lisa T. Weinbrenner \and Otfried Gühne}
\shortauthor{L. T. Weinbrenner, O. Gühne}

\institute{                    
  Naturwissenschaftlich-Technische Fakult\"at, Universit\"at Siegen, Walter-Flex-Stra\ss{}e 3, 57068 Siegen, Germany
}

\abstract{
Quantum entanglement between several particles is essential 
for applications like quantum metrology or quantum cryptography, 
but it is also central for foundational phenomena like quantum 
non-locality. This leads to the problem of quantifying
the amount of entanglement in a quantum state. We present
a review on the geometric measure of entanglement, being a quantifier 
based on the distance of a state to the nearest separable state. 
We explain basic properties, existing methods 
to compute it, its operational interpretations, as well as 
scaling and complexity issues. We point out intimate relations 
to fundamental problems in mathematics concerning  
eigenvalues and norms of tensors. Consequently, the geometric
measure of entanglement provides a playground where physical intuition 
and mathematical rigor benefit from each other. 
}

\begin{document}

\maketitle

\section{Introduction}
Entanglement plays an important role in quantum information theory
and is often considered to be a resource for quantum 
metrology, quantum cryptography or other applications 
\cite{Horodecki2009, GuhneToth2009}. As such, there is 
an ongoing search for characterization and quantification 
techniques, measuring the amount and usefulness of entanglement 
in quantum states \cite{PlenioVirmani2007, EltschkaSiewert2014}. 
Here, different use cases 
lead to different approaches on entanglement quantification, 
resulting in different measures of entanglement. 
One approach for quantifying entanglement is given by 
the geometric measure of entanglement, which quantifies the entanglement 
of a state by its distance to the separable states. 
This entanglement measure was first introduced by Shimony 
for bipartite pure states \cite{Shimony1995}, extended to 
multipartite pure states by Barnum and Linden
\cite{BarnumLinden2001} and fully generalized to mixed states 
by Wei and Goldbart \cite{WeiGoldbart2003}. 

A remarkable and important feature of the geometric measure is its close
relation to different concepts from mathematics. Not only is the 
geometric measure directly linked to the injective tensor norm, it 
is also related to the theory of tensor eigenvalues and tensor 
decompositions. In this article we wish to point out different 
connections between the physical and mathematical notions, hopefully 
leading to a better understanding of the interplay of physics and 
mathematics.

\section{Basic definitions}

We start with the simple case of a pure state $\ket{\psi} = \sum_{ij} \tau_{ij} \ket{i}_A \otimes \ket{j}_B$ of two particles, each described as a $d$-dimensional system. Here, 
the complex coefficients $\tau_{ij}$ form a $d\times d$ 
matrix. If this state can be written as 
\begin{equation}\label{eq:defSepBipartitePureState}
\ket{\psi} = \ket{a}_A \otimes \ket{b}_B \equiv \ket{ab},
\end{equation}
each subsystem can independently be described by the 
respective $d$-dimensional vectors $\ket{a}_A$ and 
$\ket{b}_B$. Consequently, the state $\ket{\psi}$ is 
called separable. If this is not the 
case, the state $\ket{\psi}$ is called entangled.

For checking the separability of $\ket{\psi}$ note 
that every state can be written in the Schmidt 
decomposition 
\begin{equation}
\label{eq:defSchmidt}
\ket{\psi} = \sum_{i=1}^r s_i \ket{a_i b_i}
\end{equation}
where the real  {Schmidt coefficients} $s_i$
are positive, unique, fulfill the normalization 
$\sum_{i=1}^r s_i^2 = 1$ and are typically ordered 
as $s_1 \geq s_2 \geq \dots$. The  {Schmidt 
vectors} $\ket{a_i}$ (and $\ket{b_j}$) 
are orthonormal and are also unique, if the Schmidt
coefficients are pairwise different.
The number $r$ is the so-called  {Schmidt rank} of the state. 
The Schmidt  decomposition follows directly from the singular 
value decomposition of the complex coefficient matrix 
$\tau_{ij}$. Clearly, a state is separable, if and only if 
the Schmidt rank equals one, $r=1$; or, in other words,
if the matrix $\tau_{ij}$ has rank one. 
The states with maximal Schmidt rank $r=d$ and identical 
Schmidt coefficients $s_i=\nicefrac{1}{\sqrt{d}}$ are 
the generalized Bell states and  maximally 
entangled.

Given an entangled state, one can naturally consider its 
distance to the set of all separable states as a quantifier 
of the amount of entanglement. In two-particle systems 
this maximal overlap $\Lambda(\psi)$ can be easily calculated 
by using  the Schmidt decomposition, resulting in
\begin{equation}
\label{eq:resultBipartitePure}
\Lambda({\psi}) :=  
\max_{\ket{a b}} | \langle{a b}\ket{\psi}| = s_1.
\end{equation}
In the mathematical literature, this is known as the 
variational characterization of the Ky-Fan norm of 
$\tau_{ij}$ \cite{HornJohnson1994}. One then defines the 
{geometric measure of entanglement} \cite{Shimony1995, BarnumLinden2001, WeiGoldbart2003} 
\begin{equation}
\label{eq:defGeomMeasure}
E({\psi}) = 1- \Lambda^2(\psi).
\end{equation}
Clearly, this measure vanishes for all separable states, 
while the maximum is given by $E=1-\nicefrac{1}{d}$. 

Naturally, one would like to apply this approach also to 
multipartite states. We consider as an example 
a tripartite state 
$
\ket{\psi} = \sum_{ijk} \tau_{ijk} \ket{ijk},
$
but the following definitions can directly be generalized 
to $N$ parties. The maximal overlap for three parties is,
as in the bipartite case,
\begin{equation}
\label{eq:overlapTripartitePure}
\Lambda({\psi}) =  \max_{\ket{abc}} 
\vert \langle{a b c}\ket{\psi}\vert
\end{equation}
and the geometric measure  $E(\psi)= 1-\Lambda^2(\psi)$.

In practice, however, a significant difference from the bipartite
case arises: The coefficients are given by a tensor $\tau_{ijk}$, 
and no easy-to-calculate singular value decomposition for 
tensors exists. Consequently, there is no Schmidt decomposition 
in the multipartite case (see also the discussion after 
Eq.~(\ref{eq-multipartite-schmidt}) below), complicating the 
calculation of the geometric measure considerably. 

As examples, consider for three qubits the 
Greenberger-Horne-Zeilinger (GHZ) state and the W state,
\begin{align}
\ket{GHZ} &= \frac{1}{\sqrt{2}} \left( \ket{000} + \ket{111} \right),
\label{eq:defGHZstate}
\\
\ket{W} & =  \frac{1}{\sqrt{3}} \left( \ket{001} + \ket{010} + \ket{100} \right). 
\label{eq:defWstate}
\end{align}
For the GHZ state, we can resort to the bipartite case by noticing that the state is already in the Schmidt 
decomposition for the bipartition $A|BC$ with Schmidt coefficients $s_1=s_2=\nicefrac{1}{\sqrt{2}}$. 
Consequently, the maximal overlap with biseparable states
of the type $\ket{\eta}=\ket{a}_A \otimes \ket{\phi}_{BC}$ is given by $\nicefrac{1}{\sqrt{2}}$. Since a product state $\ket{abc}$
is a special case of the state $\ket{\eta}$, this is also
an upper bound on $\Lambda (GHZ)$. In fact, the product 
state $\ket{000}$ reaches exactly this value, leading 
to $\Lambda^2({GHZ}) = \nicefrac{1}{2}$ and 
$E({GHZ})=\nicefrac{1}{2}$.

For the W state, this method gives only an upper bound. But, 
as we will explain in the next section, we can use the fact 
that for a permutationally symmetric state the maximizing 
product state is also symmetric, i.e. one can restrict the attention to product states of the form $\ket{aaa}$. 
Using this, one finds $\Lambda^2({W}) = \nicefrac{4}{9}$ and $E({W}) = \nicefrac{5}{9}$, so for the geometric measure, 
the W state is more entangled than the GHZ state.

\section{Basic results}

Let us start by discussing some methods to calculate or 
estimate the geometric measure. First, a simple numerical 
approach to calculate $\Lambda(\psi)$ can be formulated as a
see-saw algorithm  \cite{GuhneReimpellWerner2007, 
GerkeVogelSperling2018}. 
This relies on the fact that if 
all but one part of the product state are fixed, then the 
optimal remaining part can directly be found. More precisely, 
for given fixed states $\ket{b_0}$ and $\ket{c_0}$ one can calculate 
the unnormalized state $\ket{\alpha}=\braket{b_0 c_0}{\psi}$ on 
Alice's space. Then, it is clear that the overlap  
$|\braket{a b_0 c_0}{\psi}|$ is maximal, if we choose
$\ket{a}$ proportional  to  $\ket{\alpha}$, that is 
$\ket{a} \propto \ket{\alpha}.$ This can be extended to an
iteration, where one starts with a random product state
$\ket{a_0b_0c_0}$ and then updates first Alice's, then Bob's
and then Charlie's vector, and then starts with Alice again, 
etc.
Repeating this leads to a monotonically increasing sequence of 
values and hence to a fixed point. This is often (but not always) 
the global optimum, in practice it works well for states up to 
about seven qubits \cite{GuhneReimpellWerner2007, 
GerkeVogelSperling2018, SteinbergGuhne2022}. In fact, the fixed points are the so-called singular values of the tensor $\tau_{ijk}$, which are related to Z-eigenvalues of tensors, see also below. 

Strictly speaking, the iteration above provides only a lower 
bound on the value of $\Lambda(\psi).$ Upper bounds can be derived considering a relaxation of the problem using the
separability criterion of the positivity of the partial 
transposition (PPT) \cite{Peres1996, Horodecki1996}. For that, note that
instead of optimizing the overlap with all pure separable 
states, one can define alternatively 
$\Lambda^2(\psi) = \max_{\sigma  \in \text{f.s.}} 
\Tr(\ketbra{\psi}{\psi} \sigma)$
where the optimization runs over all fully separable 
states of the form $\sigma = \sum_k p_k \ketbra{a_k b_k c_k}{a_k b_k c_k}$ \cite{GuhneToth2009}. The set of fully separable states is strictly smaller 
than the set  of states obeying the
PPT criterion for each bipartition, that is, 
$F_{PPT} = \{\rho, \rho^{T_X} \geq 0, X=A,B,C \}$. 
So, 
\begin{equation}
    \Lambda^2(\psi)
    \leq \max_{\rho \in F_{PPT}} \Tr(\ketbra{\psi}{\psi}\rho)
    \label{eq-pptbound}
\end{equation}
denotes an upper bound. Importantly, the optimization over fully PPT states 
can be formulated as a semidefinite program, which can be solved efficiently, 
see Ref.~\cite{WeinbrennerBaksovaDenkerMorelliYuFriisGuhne2024}
for a recent concrete example. The bound (\ref{eq-pptbound}) is better 
than the upper bound from considering all bipartite Schmidt decompositions 
of the state $\ket{\psi}$ (as outlined above for the GHZ state), which follows
from connections between quantum state fidelities and the PPT criterion \cite{PianiMora2007,GuhneMaoYu2021}.

Often the computation of the geometric measure can be simplified
if $\ket{\psi}$ has specific properties. As a simple 
example, if $\ket{\psi}$ has only real and positive coefficients, 
then the product state can be assumed to have the same structure,
see, e.g., \cite{NollerGuhneGachechiladze2023}.
A highly relevant case arises if the state 
$\ket{\psi}$ is symmetric, i.e., if the state is invariant 
under the permutation $P_{\alpha, \beta}$ of any two particles 
$\alpha$ and $\beta$,
$
\ket{\psi} = P_{\alpha, \beta} \ket{\psi}.
$
Symmetric states of $N$ parties of dimension $d$ form a subspace of 
dimension ${N+d-1}\choose{d-1}$ and the entanglement theory in this 
space is well developed, see, e.g., Refs.~\cite{TothGuhne2009,WolfeYelin2014,MarconiAloyTuraSanpera2021,MarconiImai2025}.

For example, the GHZ and the W state are symmetric states. It seems rather natural that the maximal overlap $\Lambda(\psi)$ with a symmetric state should be attainable by a symmetric product 
state, that is, a state of the form $\ket{aaa}.$ Indeed, this 
was known already in mathematics for some time in the context 
of symmetric polynomials over Banach spaces \cite{Hormander1954},
see also the discussion in \cite{HubenerKleinmannWeiGonzalezGuhne2009}. 
In fact this result was then extended by proving that, in the case of 
$N\geq 3$ parties, the closest product state 
is necessarily symmetric \cite{HubenerKleinmannWeiGonzalezGuhne2009}. 
Note, however, that not every structure of the state $\ket{\psi}$ can 
directly be also assumed for the closest product state. A simple example 
is the translationally invariant state 
$
\ket{\psi} = (\ket{0101} + \ket{1010}) / \sqrt{2},
$
for which $\Lambda^2(\psi) = \nicefrac{1}{2}$  
is attained for the product states $\ket{0101}$ and $\ket{1010}$, 
which are not translationally invariant
\cite{HubenerKleinmannWeiGonzalezGuhne2009}.

A direct consequence of these results is that for the W state
$\Lambda^2(W) = \nicefrac{4}{9}$, as explained above.
In addition, the geometric measure is also known for other families 
of states, such as Dicke states and other symmetric states \cite{WeiGoldbart2003, ChenXuZhu2010,MartinGiraudBraunBraunBastin2010},
or graph and hypergraph states \cite{MarkhamMiyakeVirmani2007,HajdusekMurao2013,NollerGuhneGachechiladze2023}.

The discussed methods can be used to obtain structural insights into the geometric measure. First, the see-saw algorithm can be extended (and used 
as a subroutine) to search for maximally entangled states \cite{SteinbergGuhne2022}. The main idea is the following: 
Consider a state $\ket{\psi}$ and the closest product state 
$\ket{abc}$, potentially computed by the see-saw algorithm,
and the updated state 
$\ket{\phi} = (\ket{\psi} + \varepsilon \ket{\eta})/ \mathcal{N}$,
where $\ket{\eta} \propto [\mathbf{1} -\ketbra{abc}{abc}] \ket{\psi}$ is chosen orthogonal to $\ket{abc}$ and 
$\mathcal{N}$ denotes a normalization. It can 
be shown that if the closest product state $\ket{abc}$ is 
unique and $\varepsilon$ is sufficiently small, then
the geometric measure of $\ket{\phi}$ is strictly larger
than the one of $\ket{\psi}.$
Iterating this procedure leads to states 
with maximal geometric measure, 
for example, for three qubits one finds the W state. 
This works well up to six qubits, but also in higher dimensions \cite{SteinbergGuhne2022}.

A variation of this algorithm can be used to search for highly entangled subspaces. 
Interestingly, for three qubits one finds a two-dimensional subspace 
spanned by the W state in Eq.~(\ref{eq:defWstate}) and 
\begin{equation}
\label{eq:defVstate}
\ket{V} = \frac{1}{\sqrt{3}} \left( \ket{011} + e^{\frac{2\pi}{3}i}\ket{110} + e^{\frac{4\pi}{3}i}\ket{101} \right) .
\end{equation}
All states in this subspace have the maximally reachable geometric measure for three qubits of $\nicefrac{5}{9}$, similar constructions can be generalized \cite{DenkerImaiGuhne2025}.

The results about symmetric pure states, moreover, can be 
used to show the non-multiplicativity of $\Lambda(\psi)$, 
often also called non-additivity, if one considers the 
logarithm $\hat E(\psi) = -\log[\Lambda(\psi)]$ as a 
variant of the geometric measure. 
For that, one takes two $N$-particle states
$\ket{\psi}$ and $\ket{\phi}$ and considers the combined
state $\ket{\psi} \otimes \ket{\phi}$ as an $N$-partite 
state, but with higher local dimension. Then, in general 
it does not hold that $\Lambda(\psi \otimes \phi) = \Lambda(\psi)\Lambda(\phi)$. 
Indeed, considering a 
completely antisymmetric state $\ket{\psi}_{\rm as}$ 
(meaning that $P_{\alpha, \beta} \ket{\psi} = -\ket{\psi}$ 
for all permutations) we find that 
the closest product state cannot be symmetric, as otherwise 
their overlap would vanish. However, the state 
$\ket{\psi}_{\rm as} \otimes \ket{\psi}_{\rm as}$ is 
{\it symmetric} on the $N$-partite system, so the 
nearest product state in this higher-dimensional space 
{\it has} to be symmetric and can therefore not consist of
two copies of the closest product state to 
$\ket{\psi}_{\rm as}$ \cite{WernerHolevo2002,ZhuChenHayashi2010}.
This non-additivity is related to the
additivity of the output purity of quantum channels \cite{WernerHolevo2002,HaydenWinter2008,DilleyChangLarsonChitambar}, see also below.

\section{Operational interpretations}

The geometric measure  has 
an operational interpretation in the context 
of state discrimination. Considering $K$ orthogonal states 
$\{\ket{\psi_i} , i=1, \dots, K\}$
one can ask whether one 
can discriminate them perfectly. While this is
clearly possible if global 
measurements are allowed, the question is more 
interesting
and more relevant
if one allows only local operations and classical 
communication (LOCC).
In fact, the maximal number of states 
that can be discriminated by LOCC 
is upper bounded by their geometric measure 
\cite{HayashiMarkhamMuraoOwariVirmani2006}.

To see this, note that perfect state discrimination 
of the states $\{\ket{\psi_i}\}$ is only 
possible if there exists a generalized 
measurement with the effects $\{M_i\}$, 
such that each effect detects one of the 
states perfectly, that is 
$\Tr(M_i \ket{\psi_i}\!\bra{\psi_i}) = 1$. 
If the parties are only allowed to perform 
LOCC operations, the effects $M_i$ 
additionally have to be separable. Noting 
that the effects correspond to
density matrices $\omega_i$ up to some scaling,  optimizing the 
measurement for the given state set reduces 
to an optimization over separable density 
matrices, similar to the optimization
over product states in the
geometric measure. More precisely, one finds 
that if the discrimination of the states 
$\{\ket{\psi_i} , i=1, \dots,K\}$ 
in a $d^N$-dimensional space is possible, then 
$
\sum_{i=1}^{K} \nicefrac{1}{\Lambda^2({\psi_i})} \leq 
d^N
$ holds.
Since separable states have $\Lambda^2({\psi_i})=1$, one finds that $d^N$ separable states forming a basis can potentially be discriminated (although this does not need to be possible  \cite{BennettDiVincenzoFuchsMorRainsShorSmolinWootters1999}), but choosing one of the states to be entangled (i.e. $\Lambda^2({\psi_i}) < 1$), perfect discrimination by LOCC is no longer possible. More generally, the more entangled the states are
in terms of the geometric measure, the fewer states can be perfectly discriminated.

There are further operational interpretations, e.g., in the
context of measurement-based quantum computation or in relation 
to quantum channels, which we will explain below. 
Finally, the geometric measure has been used to characterize
phase transitions in condensed matter systems, see 
Refs.~\cite{WeiDasMukhopadyayVishveshwaraGoldbart2005,OrusWei2010,
ZhangWeiLaflamme2011,
BuerschaperGarciaSaezOrusWei2014,ShiWangLiChoBatchelorZhou2016} for examples.

\section{Asymptotic behaviour}

Having reviewed properties of the geometric measure for specific 
states, the question arises whether statistical statements 
for general states can be made, especially 
for many particles. 

A simple argument supports the conjecture that generic states
of many particles are highly entangled: The description 
of a generic state of $N$ particles requires an exponentially 
increasing number of parameters (in $N$), while for product
states this number increases only linearly. So, the fraction 
of product states decreases with $N$ and a generic state can 
be expected to be highly entangled. In fact, it was already 
shown in Ref.~\cite{HaydenLeungWinter2006} that for a generic 
multiparticle state (distributed according to the Haar measure)
the probability that there is {\it some} bipartite cut, for which 
the state is {\it not} highly entangled (measured by the entropy 
of the reduced state) is exponentially suppressed.

Concerning the geometric measure for multiqubit
states, it was shown \cite{GrossFlammiaEisert2009} that for 
$N>11$ the probability of finding a state with small geometric
measure is constrained via 
\begin{equation}
\mbox{Prob}\{\Lambda^2(\psi) > 3 N^2 \times 2^{-N} \} 
< \exp\{-N^2\}.
\label{eq-grossbound}
\end{equation}
Note that expressing a normalized state in the 
computational basis yields for any state the 
bound $\Lambda^2(\psi) \geq 2^{-N}$, so Eq.~(\ref{eq-grossbound})
implies that generic multi-qubit states have a geometric measure close to the maximally possible one.  

The proof idea is based on two facts.  First, it is well 
known that if one takes a fixed vector $\ket{\phi}$ and 
a random vector $\ket{\psi}$ in a high-dimensional space, 
then their overlap is with high probability close to zero. 
Second, one can approximate product states with a 
so-called $\varepsilon$-net. This is a small set of states 
$\{\ket{\phi_i}\}$, such that for any product state 
$\ket{\pi}$ one finds a state in the $\varepsilon$-net 
which approximates $\ket{\pi}$ well. The $\varepsilon$-net 
for $N$ qubits can be constructed from approximations 
of the Bloch sphere for a single qubit.

These facts can be combined as follows. For a random state 
$\ket{\psi}$ one asks for the probability that the maximum
of the overlaps with states in the $\varepsilon$-net is high.
Using a union bound, this can be upper bounded by the product 
of the number of states in the $\varepsilon$-net and the probability 
for a fixed state $\ket{\phi_i}$ to have a high overlap with $\ket{\psi}$.
Due to the first fact, the latter is 
also bounded, leading to the conclusion 
that the probability of $\ket{\psi}$ having a good approximation in the
$\varepsilon$-net is small. Since the $\varepsilon$-net 
is a good approximation of the product states, this also
holds if one wishes to find the maximal overlap of 
$\ket{\psi}$ with product states.

This fact has direct consequences for the usefulness of generic
states in measurement-based quantum computation \cite{BriegelBrowneDurRaussendorfNest2009}, where one prepares an entangled 
state on many qubits, for instance the so-called cluster state, and then implements an arbitrary quantum algorithm by performing local measurements. 
Naturally, this requires strong
correlations between the outcomes of the single qubits. The
result in Eq.~(\ref{eq-grossbound}) implies, however, that
for a generic state the measurement results in any local basis
are highly uncorrelated \cite{GrossFlammiaEisert2009, BremnerMoraWinter2009}. Hence, generic states cannot
be used for measurement-based quantum computation or, in the words of Ref.~\cite{GrossFlammiaEisert2009}, they are {\it too entangled to be useful}.

We add that further scaling results of the geometric measure have been derived, for instance if states have a permutational symmetry \cite{FriedlandKemp2018}, for higher dimensions \cite{ZhuChenHayashi2010}, as well as upper bounds 
obtainable for fixed dimensions \cite{DerksenFriedlandLimWang2017}.
An interesting open question is whether there is a constant $C$ such that for any $N$ there is a multi-qubit state for which $\Lambda^2(\psi) \leq C \times 2^{-N}$ (see Problem 8.27 in Ref.~\cite{AubrunSzarek2017}). More discussions, also about the 
distribution of $\Lambda^2(\psi)$ for small $N$ can be found in
Refs.~\cite{SteinbergGuhne2022,FitterLancienNechita2022}.

\section{Mathematical concepts}
Here we discuss related concepts
from multilinear algebra and tensor analysis.
\subsection{Injective tensor norm}
First, the quantity
$
\Vert \tau \Vert = 
\sup_{
\Vert{x}\Vert =
\Vert{y}\Vert =
\Vert{z}\Vert=1}
|\sum_{ijk} \tau_{ijk} x_i y_j z_k|
$, 
equaling $\Lambda(\psi)$ in our notation, is also called the 
{injective tensor norm}, and several other problems have 
been shown to be related to its computation. For instance, 
considering a quantum channel $S(\rho) = \sum_k A_k \rho A_k^\dagger$
in the Kraus representation, one~may be interested in the maximal 
purity of the output of
this channel. For that, one can consider the quantity 
$\mathcal{P}=\bra{\eta} S(\ketbra{\phi}{\phi})\ket{\eta}$,
which can be calculated as  $\mathcal{P}= \sum_k |\bra{\eta} A_k \ket{\phi}|^2 = \sup_{\Vert{x}\Vert =1} 
|\sum_k x_k\bra{\eta} A_k \ket{\phi}|^2 =
\sup_{\Vert{x}\Vert =1} |\sum_{ijk} x_k \eta^*_i A_k^{ij} \phi_j|^2$,
where the indices $i,j$ label the entries of the matrices $A_k.$
Thus, $\sup_{\ket{\eta}, \ket{\phi}} \mathcal{P} = 
\Vert A_k^{ij} \Vert^2$ holds, 
relating the additivity of the output purity 
of channels 
to the additivity of the injective tensor norm~\cite{WernerHolevo2002}.

Moreover, the complexity of computing the injective tensor norm can be 
directly related to the complexity of the bipartite separability problem 
for mixed states. For that, note that the basic idea of the iterative 
algorithm for computing $\Lambda^2(\psi)$ can be rephrased as follows: If we 
define the bipartite operator $X_{AB} = \Tr_C(\ketbra{\psi}{\psi})$ on 
the first two subsystems, then
\begin{equation}
\Lambda^2(\psi) = \sup_{\ket{a}, \ket{b}} 
\bra{ab} X_{AB} \ket{ab}.
\label{eq-bipartite}
\end{equation}
In other words, the maximal overlap of a bipartite operator
$X_{AB}$ with product states equals the parameter $\Lambda^2(\psi)$
of its purification. The optimization in Eq.~(\ref{eq-bipartite})
is essentially equivalent to the construction of bipartite
entanglement witnesses, allowing to translate 
complexity results from the bipartite separability problem
to the injective tensor norm on higher order tensors, 
see Ref.~\cite{HarrowMontanaro2013} for a detailed discussion.
Finally, tensors with maximal injective tensor norm are cases where the solution of the MAX-N-local Hamiltonian problem, where one asks for the eigenvector to the largest eigenvalue of a Hamiltonian, and its best product approximation deviate maximally \cite{GharibianKempe2012}.

\subsection{Tensor eigenvalues}
Another related concept are eigenvalues of tensors. For their
study, typically real and symmetric tensors are considered \cite{Qi2005}.
The reason is that one is interested in the positivity of polynomials 
of the type
$f(x) = \sum \tau_{ijkl} x_i x_j x_k x_l = 
\langle{\psi}\ket{xxxx}$
for real tensors $\tau_{ijkl}$, and such a polynomial does not capture non-symmetric properties of the tensor. Additionally, for a given non-symmetric tensor one can also always consider a symmetric embedding into a larger symmetric tensor \cite{RagnarssonVanloan2013}. We here focus our discussion on tensors with four indices, but anything can directly be generalized to $N$ indices.

For real symmetric tensors, there are two important classes of 
tensor eigenvalues. The first class are the {H-eigenvalues}, which 
are defined by
\begin{equation}
\langle{\psi}\ket{xxx} = \lambda \bra{x^{3}} \mbox{ $\Leftrightarrow$ }
\sum_{jkl} \tau_{ijkl} x_j x_k x_l = \lambda x_i^3
\end{equation}
where $\lambda \in \mathbb{R}$ is the real eigenvalue, $\ket{x} \neq 0$ 
is the real eigenvector and $\ket{x^{3}}$ denotes the vector 
with the components $x_i^{3}$. The seemingly strange power of 
$3$ (or, more generally, $N-1$) on the RHS assures for a given 
$\lambda$ that every (scaled) vector $\alpha\ket{x}$ is an 
eigenvector, too. The H-eigenvalues fulfill many of the 
properties of matrix eigenvalues. For instance, the H-eigenvalues 
of a real tensor for even $N$ are exactly the diagonal elements, 
there exists a Gershgorin-disc-type theorem and the product of 
the eigenvalues equals $\det(\tau)$, where $\det(\cdot)$ is 
the hyperdeterminant used to calculate the eigenvalues 
\cite{Qi2005}. 

The second class are {Z-eigenvalues}, defined by
\begin{equation}
    \langle{\psi}\ket{xxx} = 
    \lambda \bra{x} \mbox{ with } \braket{x}{x} = 1,
\label{eq-z-eigenvalue}
\end{equation}
where again $\lambda \in \mathbb{R}$ is the real eigenvalue and 
$\ket{x}\neq 0$ is the corresponding eigenvector. These eigenvalues 
are~not directly comparable to matrix eigenvalues, as scaling a 
Z-eigenvector $\alpha\ket{x}$ 
leads in general not to an Z-eigenvector 
again. Additionally, the Z-eigenvalues are not necessarily given 
by the diagonal elements of diagonal tensors. As an example, 
consider the (unnormalized) state 
$
\ket{\psi} = 3 \ket{0000} + \ket{1111},
$
where the coefficient tensor $\tau$ is diagonal with diagonal elements $\tau_{0000}=3$ and $\tau_{1111}=1$. Besides the two obvious Z-eigenpairs 
$\ket{x^{(1)}}=\ket{0}$ for $\lambda^{(1)}=3$, and $\ket{x^{(2)}} = \ket{1}$ 
for $\lambda^{(2)}=1$, there exists the third Z-eigenpair $\ket{x^{(3)}}=\nicefrac{1}{2}(\ket{0} + \sqrt{3}\ket{1})$ for 
$\lambda^{(3)}=\nicefrac{3}{4}$ \cite{Qi2005}. This property is  
important when calculating the eigenvectors or the geometric measure
by a see-saw algorithm, as every eigenvector is a fixed point of the 
iteration. On the other hand, Z-eigenvalues have the nice property of 
being invariant under any local orthogonal transformation 
$\ket{\psi} \mapsto O \otimes O \otimes O \ket{\psi}$. Both classes of eigenvalues 
are related to the positivity of the polynomial $f(x)$ for even $N$, 
as the polynomial is positive, if and only if all H-eigenvalues (or, equivalently, all Z-eigenvalues) are positive \cite{Qi2005}.

The above notions can be generalized to complex tensors, complex 
eigenvalues or complex eigenvectors, leading to eigenvalues denoted 
by different letters. For example, {E-eigenvalues} can be obtained 
from the definition of Z-eigenvalues as in Eq.~(\ref{eq-z-eigenvalue}) 
by allowing for complex eigenvalues $\lambda\in\mathbb{C}$ and 
eigenvectors $\ket{x}\in\mathbb{C}^d$, but still considering the 
euclidean normalization condition $ \sum_i x_i^2 = \braket{x^*}{x} = 1$.
This is not the optimization leading to $\Lambda^2(\psi)$ for 
complex quantum states, as the normalization condition is relevant. 
For example, for
$\ket{\psi} = \ket{000000} + \ket{111111}$, the state 
$\ket{x}=\nicefrac{1}{\sqrt{2}}(\ket{0} + i\ket{1})$ fulfills 
the first condition for an E-eigenvalue, but its euclidean norm is
$\sum_i x_i^2 = \braket{x^*}{x} = 0$, so it is not normalizable. 

Another relevant class are the  {US-eigenvalues} of complex symmetric 
tensors, which are directly related to the geometric measure \cite{NiQiBai2014}. These are defined as in Eq.~(\ref{eq-z-eigenvalue}) 
where $\lambda$ is the real eigenvalue and $\ket{x}$ is the 
according (complex) eigenvector with the standard normalization 
$\sum_i |x_i|^2 = \braket{x}{x} = 1$. This definition directly 
yields the geometric measure of a symmetric state as the maximal 
US-eigenvalue. If the state is in addition  real and non-negative,
the maximal US-eigenvalue reduces to the largest Z-eigenvalue
\cite{HuQiZhang}.

Concerning the computation of eigenvalues, we would like to add two 
more facts. First, for finding the largest US-eigenvalue one may 
consider the set $\mathcal{X} = \{ \Vert x \Vert \mid \bra{\psi} (\ket{x}^{\otimes (N-1)}) 
= \bra{x}, \ \ket{x} \neq 0 \}$. Then, $\max\{ \lambda \} = (\nicefrac{1}{\min \mathcal{X}})^{N-2}$, transforming the problem into a system of algebraic 
equations \cite{NiQiBai2014}. 
A different approach to calculate the different eigenvalues is  the so-called 
tensor power method, being similar to the see-saw algorithm described above. In this method, one starts from a symmetric state $\ket{x^{(0)}x^{(0)}x^{(0)}x^{(0)}}$, and, after calculating the overlap $\langle{\psi}\ket{x^{(0)}x^{(0)}x^{(0)}}$, updates every part of the vector according to $\ket{x^{(1)}} \propto \langle{\psi}\ket{x^{(0)}x^{(0)}x^{(0)}}$.

Finally, one can also define singular values 
for tensors.  For that, one considers an entire set $\{\ket{x^{(1)}}, 
\dots, \ket{x^{(N)}}\}$ of vectors which obey equations like
\begin{equation}
    \braket{\psi}{x^{(1)},\dots,\bar{x}^{(k)},\dots , x^{(N)}} = \lambda \bra{x^{(k)}},  \quad
    \braket{x^{(i)}}{x^{(i)}} = 1 \; \forall i.
\end{equation}
The notation $\bar{x}^{(k)}$ denotes that the vector $\ket{x^{(k)}}$ is missing in the tensor product. Naturally, this is closely related to the notion of
entanglement eigenvalue equations, which have been used to analyze the geometric measure \cite{WeiGoldbart2003, GerkeVogelSperling2018}.

We would like to add that tensor eigenvalues have indeed been
used in quantum physics besides the notions of the geometric measure.
In Ref.~\cite{BohnetWaldraffBraunGiraud2016} the authors expressed symmetric multi-qubit states in the Bloch representation, i.e., in terms of expectation values of tensor
products of Pauli matrices. This leads to a real, symmetric
tensor, and it is then shown that if the smallest Z-eigenvalue is negative, the state is entangled.

\subsection{Odeco and Fradeco decompositions}

The concepts of the geometric measure, the injective tensor norm as well as 
tensor eigenvalues aim at approximating a tensor $\tau_{ijk}$ or 
specific properties of it by a tensor of the form $\eta_{ijk}=a_i b_j c_k$,
which may be considered as a rank-one tensor. It is natural to ask for further mathematical concepts to define 
extended notions of correlations in quantum information.

A first question concerns the generalization of the Schmidt 
decomposition to three or more particles, asking 
whether a state can be written as
\begin{equation}
\ket{\psi} = \sum_{i=1}^r s_i \ket{a_i b_i c_i},
\label{eq-multipartite-schmidt}
\end{equation}
where the  $\ket{a_i}$ ($\ket{b_i}$, $\ket{c_i}$) form an orthonormal
basis of the respective local Hilbert space. A simple argument shows that
this is in general not possible \cite{Peres1995}: A decomposition as in
Eq.~(\ref{eq-multipartite-schmidt}) constitutes also a bipartite Schmidt
decomposition as in Eq.~(\ref{eq:defSchmidt}) for the partition~$A|BC$, but with the additional constraint that the Schmidt vectors on 
the joint party $BC$ are product vectors. Since the bipartite Schmidt 
decomposition in  Eq.~(\ref{eq:defSchmidt}) is unique, one cannot expect
this condition to be met for generic~$\ket{\psi}$. Still, normal forms 
of multiparticle states have been derived \cite{AcinAndrianovCostaJaneLatorreTarrach2000,CarteretHiguchiSudbery2000}.

A decomposition as in Eq.~(\ref{eq-multipartite-schmidt}) 
is possible for a specific class of tensors, and in the mathematical 
literature real and symmetric tensors with this property are called
orthogonally decomposable (odeco) \cite{Robeva2016}. For these 
tensors, an explicit form of all Z-eigenvectors in terms of the 
Schmidt vectors in Eq.~(\ref{eq-multipartite-schmidt}) can be found 
\cite{Robeva2016}. Moreover, for odeco tensors the Schmidt vectors  are exactly the robust fixed points of the tensor power iteration outlined above, so 
that they can be found numerically \cite{Robeva2016,AnandkumarGeHsuKakadeTelgarsky2014}.

A potential generalization of odeco tensors are tensors that can 
be decomposed into frames (fradeco), 
i.e., tensors where a 
decomposition as in 
Eq.~(\ref{eq-multipartite-schmidt}) can be found and the  
$\ket{a_i}$ (and $\ket{b_i}$ and $\ket{c_i}$) form a frame instead of a basis
\cite{OedingRobevaSturmfels2016}.
A (unit-norm) frame is a set of $r$ (normalized) vectors $\{ \ket{v_k} \} $ that 
form an overcomplete basis in the sense that for any vector 
$\ket{w}$ one has
$
A \Vert w \Vert ^2 \leq \sum_{i=1}^r |\langle{w}|{v_k}\rangle|^2 \leq B \Vert w \Vert ^2   
$
with some constants $A$ and $B$, and the frame is tight if $A=B$
(for an introduction on frames see, e.g., \cite{CasazzaLynch2016}).
For fradeco tensors, the 
tensor power method for calculating eigenvalues does not 
always find all eigenvectors, as some of them belong to
repelling fixed points \cite{CzaplinskiRaaschSteinberg2023}.

More generally, one can ask how many terms 
in a decomposition like Eq.~(\ref{eq-multipartite-schmidt})
are needed, even if no restriction on the local vectors 
$\ket{a_i}$ etc.~is made. Such decompositions can be represented graphically \cite{KlinglerNetzerDelesCoves2025}, and can be seen as dual to the injective tensor norm \cite{DerksenFriedlandLimWang2017}.
The resulting minimal $r_{\rm min}$ 
is called the tensor rank of the tensor $\tau_{ijk}.$ This is
central for many applications, such as the 
efficient numerical matrix multiplication \cite{Fawzietal2022}. Its computation, 
however, is difficult as it depends on the underlying field
(a real tensor may have a different tensor rank over $\mathbb{C}$
and $\mathbb{R}$) and it is numerically unstable (a tensor may be 
approximated arbitrarily well by tensors with strictly smaller 
rank, see  Eq.~(\ref{eq-omegadef}) below).
In quantum information theory, the tensor rank appeared as a
quantifier of entanglement, called the Schmidt measure 
\cite{EisertBriegel2001}, and as a tool to characterize
entanglement transformations \cite{ChitambarDuanShi2008} and 
correlations \cite{KlinglerNetzerDelesCoves2025}.

\section{The geometric measure for mixed states}

Up to this point, we considered the geometric measure $E({\psi})$ as a function of pure states only, but in practice, mixed quantum states are unavoidable.
To extend the definition of geometric measure to mixed states,
one needs to specify the required properties of an 
entanglement measure \cite{PlenioVirmani2007, EltschkaSiewert2014}.
Standard conditions for a measure
are that $E(\rho)$ should be positive on all states, 
i.e., $E(\rho)\geq 0$, and  $E(\rho) = 0$ should hold 
if $\rho$  is separable. Additionally, $E(\rho)$ should not increase under positive maps that can be implemented by LOCC,
so for any LOCC map $L$ one has $E(\rho) \geq E(L(\rho))$. This implies invariance of $E$ under local unitary transformations (which are reversible LOCC maps); but the class
of general LOCC maps is hard to characterize.

One possibility of extending a potential entanglement measure 
from pure states to mixed states is given by the {convex roof construction}. In this construction one considers all convex decompositions $\rho = \sum_i p_i \ket{\psi_i}\bra{\psi_i}$ of the state $\rho$ 
and takes the minimum
\begin{equation}
E(\rho) = \min_{p_i, \ket{\psi_i}} \sum_i p_i E(\psi_i)
\end{equation}
over all decompositions. Clearly, this minimization is hard 
to evaluate, and a full solution of this problem has only been achieved for specific entanglement measures, such as the two-qubit entanglement of formation \cite{Wootters1998} or 
some versions of the concurrence \cite{Uhlmann2000}.

Typically, the convex roof construction inherits nice properties
from $E(\psi)$, and for the geometric measure, it was shown that the convex roof $E(\rho)$ obeys all desired conditions
for entanglement measures \cite{WeiGoldbart2003}. Although it is not clear from the construction at all, the 
convex-roof extended geometric measure has again a geometrical
meaning \cite{StreltsovKampermannBruss2010}. If one measures
the overlap (or inverse distance) of mixed states  $\rho$ and $\sigma$  via the Uhlmann fidelity 
$F(\rho,\sigma) = [\Tr(\sqrt{ \sqrt{\rho} \sigma \sqrt{\rho} })]^2 $, then the geometric measure of mixed states quantifies the distance to the closest mixed fully separable state, via $E(\rho) = 1- \max_{\sigma \in \text{f.s.}} F(\rho,\sigma)$. 

There are some cases where the geometric measure
of mixed states can be analytically determined (e.g., for some specific bound entangled states \cite{WeiAltepeterGoldbartMunro2004}), or at least bounded from below 
\cite{TothMoroderGuhne2015,ZhangDaiDongZhang}.
For two qubits,  $E(\rho)$ is related 
to the mixed state concurrence $C(\rho)$ \cite{Wootters1998} via 
$
    E(\rho) = \frac{1}{2} \big[ 1 - \sqrt{1 - C(\rho)^2} 
    \big]
$ \cite{Vidal2000, WeiGoldbart2003} .
However, this is not easy to generalize to the multipartite
case or beyond qubits.

One can also obtain lower bounds on the geometric measure 
of an experimentally given state from incomplete measurement data, for example from measuring an entanglement witness $\mathcal{W}$ or a pure state fidelity
$F_\phi = \bra{\phi} \rho \ket{\phi}$ \cite{GuhneReimpellWerner2007,EisertBrandaoAudenaert2007}. 
For computing $\eta = \inf\{ E(\rho) \mid \Tr(\mathcal{W} \rho) = w\}$ one can use the theory of 
the Legendre transform, and for the geometric measure, this can be evaluated with a see-saw algorithm \cite{GuhneReimpellWerner2007} or even
analytically, e.g., if $\mathcal{W}$ is a fidelity measurement
\cite{GuhneToth2009}.
This has been used to calculate the geometric measure of 
mixed GHZ symmetric multi-qubit states or Dicke states 
affected by noise \cite{BuchholzMoroderGuhne2016, GuhneBodokyBlaauboer2008}.

\section{Discussion and extensions}

The geometric measure of entanglement combines physical 
relevance with mathematical elegance. It provides a 
playground where physical intuition as well as physical 
concepts (such as symmetries, separability criteria, and relaxations) help to gain insights into mathematical structures. 
Moreover, other problems such as the characterization of 
quantum channels can be related to the geometric measure, 
allowing to draw connections between different 
problems. 

For future research, there are many interesting directions
possible, here we would like to mention only two. First, 
while the quantity $\Lambda(\psi)$ characterizes the 
best approximation of $\ket{\psi}$ with product states, one 
may consider higher-order approximations, e.g., by states 
of tensor rank two, 
\begin{equation}
\Omega(\psi) = \sup_{\ket{\phi}=\ket{abc}+\ket{def}} 
|\langle \phi | \psi\rangle|.
\label{eq-omegadef}
\end{equation}
Note that this is a so-called ill-posed problem, meaning
that one has to consider the supremum instead of the 
maximum here \cite{KlinglerNetzerDelesCoves2025, deSilvaLim2008}, as the maximum may not be attained. 
This problem can be illustrated with the W state from Eq.~(\ref{eq:defWstate}), which has tensor rank three but
$\Omega(W)=1$, since it 
can be approximated arbitrarily well by 
rank two states of the form
$
 \ket{\phi}_\varepsilon \propto  
 (\ket{0} + \varepsilon \ket{1})^{\otimes 3}  
 - \ket{0}^{\otimes 3}
$
in the limit $\varepsilon\rightarrow 0$.
Other physically relevant approximations are pure stabilizer
states or states from low-dimensional manifolds of the state space, 
but for these cases not much is known.

Second, it would be highly desirable to consider the approximation $\Lambda(\psi)$ from Eq.~(\ref{eq:overlapTripartitePure}) also with other normalization constraints of $\ket{a}, \ket{b}$, and 
$\ket{c}$. Examples are additional linear constraints 
of the type $\bra{a}A\ket{a} = \alpha$. This may help to
discuss optimizations over process tensors \cite{TarantoMilzMuraoTulioModi2025}; moreover, optimizations like Eq.~(\ref{eq-omegadef})
may be reduced to Eq.~(\ref{eq:overlapTripartitePure}) with such constraints. In this way, the 
geometric measure will help
to open further perspectives on various problems in quantum information theory.

\acknowledgements
We thank Sophia Denker, Satoya Imai, Thorsten Raasch, Jonathan Steinberg, Tzu-Chieh Wei, and Xiao-Dong Yu for discussions. This work has been supported 
by the Deutsche Forschungsgemeinschaft (DFG, German Research
Foundation, project numbers 447948357 and 440958198),
the Sino-German Center for Research Promotion (Project 
M-0294), and the German Ministry of Education and Research 
(Project QuKuK, BMBF Grant No. 16KIS1618K). L.T.W. 
acknowledges support by the House of Young Talents of 
the University of Siegen.

\bibliographystyle{eplbib}
\bibliography{references}

\begin{thebibliography}{10}
\expandafter\ifx\csname url\endcsname\relax\def\url#1{\texttt{#1}}\fi

\bibitem{Horodecki2009}
\Name{Horodecki R. \etal} \REVIEW{Rev. Mod. Phys.}{81}{2009}{865}.

\bibitem{GuhneToth2009}
\Name{Gühne O. \etal} \REVIEW{Phys. Rep.}{474}{2009}{1}.

\bibitem{PlenioVirmani2007}
\Name{Plenio M.~B. \etal} \REVIEW{Quant. Inf. Comput.}{7}{2007}{1}.

\bibitem{EltschkaSiewert2014}
\Name{Eltschka C. \etal} \REVIEW{J. Phys. A: Math. Theor.}{47}{2014}{424005}.

\bibitem{Shimony1995}
\Name{Shimony A.} \REVIEW{Ann. N. Y. Acad. Sci.}{755}{1995}{675}.

\bibitem{BarnumLinden2001}
\Name{Barnum H. \etal} \REVIEW{Phys. A Math. Gen.}{34}{2001}{6787}.

\bibitem{WeiGoldbart2003}
\Name{Wei T.-C. \etal} \REVIEW{Phys. Rev. A}{68}{2003}{042307}.

\bibitem{HornJohnson1994}
\Name{Horn R.~A. \etal} \Book{Topics in matrix analysis} (Cambridge university
  press) 1994.

\bibitem{GuhneReimpellWerner2007}
\Name{G{\"u}hne O. \etal} \REVIEW{Phys. Rev. Lett.}{98}{2007}{110502}.

\bibitem{GerkeVogelSperling2018}
\Name{Gerke S. \etal} \REVIEW{Phys. Rev X}{8}{2018}{031047}.

\bibitem{SteinbergGuhne2022}
\Name{Steinberg J. \etal} \REVIEW{Phys. Rev. A}{110}{2024}{062428}.

\bibitem{Peres1996}
\Name{Peres A.} \REVIEW{Phys. Rev. Lett.}{77}{1996}{1413}.

\bibitem{Horodecki1996}
\Name{Horodecki M. \etal} \REVIEW{Phys. Lett. A}{223}{1996}{1}.

\bibitem{WeinbrennerBaksovaDenkerMorelliYuFriisGuhne2024}
\Name{Weinbrenner L.~T. \etal} \REVIEW{arXiv:2412.18331}{}{2024}{}.

\bibitem{PianiMora2007}
\Name{Piani M. \etal} \REVIEW{Phys. Rev. A}{75}{2007}{012305}.

\bibitem{GuhneMaoYu2021}
\Name{G\"uhne O. \etal} \REVIEW{Phys. Rev. Lett.}{126}{2021}{140503}.

\bibitem{NollerGuhneGachechiladze2023}
\Name{N{\"o}ller J. \etal} \REVIEW{J. Phys. A: Math. Theor.}{56}{2023}{375302}.

\bibitem{TothGuhne2009}
\Name{T\'oth G. \etal} \REVIEW{Phys. Rev. Lett.}{102}{2009}{170503}.

\bibitem{WolfeYelin2014}
\Name{Wolfe E. \etal} \REVIEW{Phys. Rev. Lett.}{112}{2014}{140402}.

\bibitem{MarconiAloyTuraSanpera2021}
\Name{Marconi C. \etal} \REVIEW{{Quantum}}{5}{2021}{561}.

\bibitem{MarconiImai2025}
\Name{Marconi C. \etal} \REVIEW{arXiv:2504.01578}{}{2025}{}.

\bibitem{Hormander1954}
\Name{H{\"o}rmander L.} \REVIEW{Math. Scand.}{2}{1954}{55}.

\bibitem{HubenerKleinmannWeiGonzalezGuhne2009}
\Name{H{\"u}bener R. \etal} \REVIEW{Phys. Rev. A}{80}{2009}{032324}.

\bibitem{ChenXuZhu2010}
\Name{Chen L. \etal} \REVIEW{Phys. Rev. A}{82}{2010}{032301}.

\bibitem{MartinGiraudBraunBraunBastin2010}
\Name{Martin J. \etal} \REVIEW{Phys. Rev. A}{81}{2010}{062347}.

\bibitem{MarkhamMiyakeVirmani2007}
\Name{Markham D. \etal} \REVIEW{New J. Phys.}{9}{2007}{194}.

\bibitem{HajdusekMurao2013}
\Name{Hajdušek M. \etal} \REVIEW{New J. Phys.}{15}{2013}{013039}.

\bibitem{DenkerImaiGuhne2025}
\Name{Denker S. \etal} \REVIEW{arXiv:2506.15609}{}{2025}{}.

\bibitem{WernerHolevo2002}
\Name{Werner R.~F. \etal} \REVIEW{J. Math. Phys.}{43}{2002}{4353}.

\bibitem{ZhuChenHayashi2010}
\Name{Zhu H. \etal} \REVIEW{New J. Phys.}{12}{2010}{083002}.

\bibitem{HaydenWinter2008}
\Name{Hayden P. \etal} \REVIEW{Commun. Math. Phys.}{284}{2008}{263}.

\bibitem{DilleyChangLarsonChitambar}
\Name{Dilley D. \etal} \REVIEW{arXiv:2503.23247}{}{2025}{}.

\bibitem{HayashiMarkhamMuraoOwariVirmani2006}
\Name{Hayashi M. \etal} \REVIEW{Phys. Rev. Lett.}{96}{2006}{040501}.

\bibitem{BennettDiVincenzoFuchsMorRainsShorSmolinWootters1999}
\Name{Bennett C.~H. \etal} \REVIEW{Phys. Rev. A}{59}{1999}{1070}.

\bibitem{WeiDasMukhopadyayVishveshwaraGoldbart2005}
\Name{Wei T.-C. \etal} \REVIEW{Phys. Rev. A}{71}{2005}{060305}.

\bibitem{OrusWei2010}
\Name{Or\'us R. \etal} \REVIEW{Phys. Rev. B}{82}{2010}{155120}.

\bibitem{ZhangWeiLaflamme2011}
\Name{Zhang J. \etal} \REVIEW{Phys. Rev. Lett.}{107}{2011}{010501}.

\bibitem{BuerschaperGarciaSaezOrusWei2014}
\Name{Buerschaper O. \etal} \REVIEW{J. Stat. Mech.: Theory
  Exp.}{2014}{2014}{P11009}.

\bibitem{ShiWangLiChoBatchelorZhou2016}
\Name{Shi Q.-Q. \etal} \REVIEW{Phys. Rev. A}{93}{2016}{062341}.

\bibitem{HaydenLeungWinter2006}
\Name{Hayden P. \etal} \REVIEW{Comm. Math. Phys.}{265}{2006}{95}.

\bibitem{GrossFlammiaEisert2009}
\Name{Gross D. \etal} \REVIEW{Phys. Rev. Lett.}{102}{2009}{190501}.

\bibitem{BriegelBrowneDurRaussendorfNest2009}
\Name{Briegel H.~J. \etal} \REVIEW{Nat. Phys.}{5}{2009}{19}.

\bibitem{BremnerMoraWinter2009}
\Name{Bremner M.~J. \etal} \REVIEW{Phys. Rev. Lett.}{102}{2009}{190502}.

\bibitem{FriedlandKemp2018}
\Name{Friedland S. \etal} \REVIEW{Proc. Am. Math. Soc.}{146}{2018}{5035}.

\bibitem{DerksenFriedlandLimWang2017}
\Name{Derksen H. \etal} \REVIEW{arXiv:1705.07160}{}{2017}{}.

\bibitem{AubrunSzarek2017}
\Name{Aubrun G. \etal} \Book{Alice and Bob meet Banach} Vol. 223 (American
  Mathematical Soc.) 2017.

\bibitem{FitterLancienNechita2022}
\Name{Fitter K. \etal} \REVIEW{arXiv:2209.11754}{}{2022}{}.

\bibitem{HarrowMontanaro2013}
\Name{Harrow A.~W. \etal} \REVIEW{J. ACM}{60}{2013}{1}.

\bibitem{GharibianKempe2012}
\Name{Gharibian S. \etal} \REVIEW{SIAM J. Comput.}{41}{2012}{1028}.

\bibitem{Qi2005}
\Name{Qi L.} \REVIEW{J. Symb. Comput.}{40}{2005}{1302}.

\bibitem{RagnarssonVanloan2013}
\Name{Ragnarsson S. \etal} \REVIEW{Linear Algebra Its Appl.}{438}{2013}{853}.

\bibitem{NiQiBai2014}
\Name{Ni G. \etal} \REVIEW{SIAM J. Matrix Anal.}{35}{2014}{73}.

\bibitem{HuQiZhang}
\Name{Hu S. \etal} \REVIEW{Phys. Rev. A}{93}{2016}{012304}.

\bibitem{BohnetWaldraffBraunGiraud2016}
\Name{Bohnet-Waldraff F. \etal} \REVIEW{Phys. Rev. A}{94}{2016}{042324}.

\bibitem{Peres1995}
\Name{Peres A.} \REVIEW{Phys. Lett. A}{202}{1995}{16}.

\bibitem{AcinAndrianovCostaJaneLatorreTarrach2000}
\Name{Ac\'{\i}n A. \etal} \REVIEW{Phys. Rev. Lett.}{85}{2000}{1560}.

\bibitem{CarteretHiguchiSudbery2000}
\Name{Carteret H.~A. \etal} \REVIEW{J. Math. Phys.}{41}{2000}{7932}.

\bibitem{Robeva2016}
\Name{Robeva E.} \REVIEW{SIAM J. Matrix Anal.}{37}{2016}{86}.

\bibitem{AnandkumarGeHsuKakadeTelgarsky2014}
\Name{Anandkumar A. \etal} \REVIEW{J. Mach. Learn. Res.}{15}{2014}{2773}.

\bibitem{OedingRobevaSturmfels2016}
\Name{Oeding L. \etal} \REVIEW{Adv. Appl. Math.}{73}{2016}{125}.

\bibitem{CasazzaLynch2016}
\Name{Casazza P.~G. \etal} \REVIEW{arXiv:1509.07347}{}{2016}{}.

\bibitem{CzaplinskiRaaschSteinberg2023}
\Name{Czapli{\'n}ski A. \etal} \REVIEW{Adv. Appl. Math.}{148}{2023}{102521}.

\bibitem{KlinglerNetzerDelesCoves2025}
\Name{Klingler A. \etal} \REVIEW{{Quantum}}{9}{2025}{1649}.

\bibitem{Fawzietal2022}
\Name{Fawzi A. \etal} \REVIEW{Nature}{610}{2022}{47}.

\bibitem{EisertBriegel2001}
\Name{Eisert J. \etal} \REVIEW{Phys. Rev. A}{64}{2001}{022306}.

\bibitem{ChitambarDuanShi2008}
\Name{Chitambar E. \etal} \REVIEW{Phys. Rev. Lett.}{101}{2008}{140502}.

\bibitem{Wootters1998}
\Name{Wootters W.~K.} \REVIEW{Phys. Rev. Lett.}{80}{1998}{2245}.

\bibitem{Uhlmann2000}
\Name{Uhlmann A.} \REVIEW{Phys. Rev. A}{62}{2000}{032307}.

\bibitem{StreltsovKampermannBruss2010}
\Name{Streltsov A. \etal} \REVIEW{New J. Phys.}{12}{2010}{123004}.

\bibitem{WeiAltepeterGoldbartMunro2004}
\Name{Wei T.-C. \etal} \REVIEW{Phys. Rev. A}{70}{2004}{022322}.

\bibitem{TothMoroderGuhne2015}
\Name{T\'oth G. \etal} \REVIEW{Phys. Rev. Lett.}{114}{2015}{160501}.

\bibitem{ZhangDaiDongZhang}
\Name{Zhang Z. \etal} \REVIEW{Sci. Rep.}{10}{2020}{12122}.

\bibitem{Vidal2000}
\Name{Vidal G.} \REVIEW{Phys. Rev. A}{62}{2000}{062315}.

\bibitem{EisertBrandaoAudenaert2007}
\Name{Eisert J. \etal} \REVIEW{New J. Phys.}{9}{2007}{46}.

\bibitem{BuchholzMoroderGuhne2016}
\Name{Buchholz L.~E. \etal} \REVIEW{Annalen der Physik}{528}{2016}{278}.

\bibitem{GuhneBodokyBlaauboer2008}
\Name{G\"uhne O. \etal} \REVIEW{Phys. Rev. A}{78}{2008}{060301}.

\bibitem{deSilvaLim2008}
\Name{de~Silva V. \etal} \REVIEW{SIAM J. Matrix Anal.}{30}{2008}{1084}.

\bibitem{TarantoMilzMuraoTulioModi2025}
\Name{Taranto P. \etal} \REVIEW{arXiv:2503.09693}{}{2025}{}.

\end{thebibliography}
    
\end{document}